\documentclass[showpacs,APS]{revtex4}

\usepackage{graphicx}
\usepackage{amsmath}
\usepackage{color}
\usepackage[normalem]{ulem}
\usepackage{multirow}

\DeclareRobustCommand\bblash{\btt{\@backslashchar}} \makeatother

\begin{document}
\title{Buchdahl compactness limit and gravitational field energy}
\author{Naresh Dadhich}
\affiliation{Inter-University Centre for Astronomy \& Astrophysics,\\ Post Bag 4, Pune, 411 007, India}
\date{\today}

\begin{abstract}
The main aim of this paper is essentially to point out that the Buchdahl compactness limit of a static object is given by \it{gravitational field energy being less than or equal to half of its non-gravitational matter energy}. It is thus entirely determined without any reference to interior distribution by the unique exterior solutions, the Schwarzschild for neutral and the Reissner-Nordstr{$\ddot o$}m for charged object. In terms of surface potential, it reads as $\Phi(R) = (M-Q^2/2R)/R \leq 4/9$ which translates to surface red-shift being less than or equal to $3$. It also prescribes an upper bound on charge an object could have, $Q^2/M^2 \leq 9/8 > 1$.
\end{abstract}


\maketitle

\section{Introduction}

Astrophysically it is a question of prime importance that how compact a static cold star like object could be, or in other words, does there exist a limit on its compactness? This question was first answered by the well known Buchdahl limit for a perfect fluid interior \cite{buch}, $
\Phi(R) = M/R\leq 4/9$ for an object of mass $M$ and radius $R$. This bound was obtained on very general conditions of density and isotropic pressure being positive and the former being non increasing outwards, $d\rho/dr \leq 0$, and the interior metric is matched at the boundary located at $p=0$, to the exterior Schwarzschild metric. This limit is given by pressure at the center $p_c = p(r=0)\leq\infty$ for the stiffest equation of state of uniform density --- incompressible perfect fluid distribution which is given by the unique Schwarzschild's interior solution not only for the Einstein but also for the Lovelock gravity in general \cite{dmk10}. Alternatively the same limit was obtained \cite{andrea08, kara-stal08} by assuming the strong energy condition, $p_r + 2p_t \leq \rho_m$, where $p_r, \, p_t, \, \rho_m$ are respectively radial and transverse pressure, and matter energy density. In this case the limit saturates not for fluid interior but for an infinitely thin shell with $2p_t = \rho_m$. In the previous case the limit saturates for central pressure tending to infinity while in the latter, it is for infinitely thin shell. In either case, situation is not entirely physically acceptable, it should anyway be taken as a limiting case. \\

There have also been various alternative derivations involving various situations like inclusion of $\Lambda$ \cite{mak-stuch}, different conditions than Buchdahl's \cite{andrea09, kara-stal08}, brane-world gravity \cite{ger-maa, gar-ure}, modified gravity theories including Lovelock gravity and higher dimensions \cite{gos, wri, sum-dad16}. The limit on maximum mass has been obtained by appealing to the dominant energy condition and sound velocity being subliminal \footnote{It should however be noted that it is sound velocity which is characteristic of fluid and it depends upon its compactness. Absolute compactness limit should however be given by incompressibility;i.e constant density fluid, which would however  violate both the preceding conditions. It is a limiting state that  is only asymptotically approachable.} \cite{bar}. The Buchdahl limit defines an overriding state which is obtained under very general conditions while more compact distributions are allowed under specific circumstances and conditions. \\

Inclusion of electric charge however brings in repulsive effect due to gravitational interaction of electrostatic energy, that would imply increase in the compactness ratio. Following the Buchdahl method under condition of matter density decreasing while charge density increasing with radius, and charged interior matching with the exterior Reissner - Nordstr{$\ddot o$}m metric at the boundary,  a compactness limit for a charged static object has been obtained \cite{roth, mak-dob-har, boe-har}. In particular, the compactness limit for a uniform density neutral distribution surrounded by a thin charged shell \cite{roth} is obtained as
\begin{equation}
M/R \leq \frac{8/9}{1 + \sqrt{1 - 8\alpha^2/9}}~, \, \alpha^2 = Q^2/M^2~.
\end{equation}
It reduces to the Buchdahl limit for uncharged object, $M/R \leq 4/9$ for $Q = 0 $, while for extremally charged, $\alpha^2 = 9/8$, non black hole object, $M/R \leq 8/9 < 1$. Note that extremality for a charged object is over-extremality $\alpha^2 >1$ for a black hole. It is interesting that $M/R < 1$ always, and more remarkably there is an upper bound on charge a static object can have, $\alpha^2 \leq 9/8$. This means non black hole object could indeed have $\alpha^2 >1$;i.e. overcharged relative to a black hole.  \\

As for uncharged fluid sphere, by employing the strong energy condition for matter distribution, $p_r + 2p_t \leq \rho_m$ and following \cite{kara-stal08}, a bound has been found \cite{andrea09} for a charged interior distribution matching with the exterior Reissner - Nordstr{$\ddot o$}m metric at the boundary, $p_r = 0$. It reads as
\begin{equation}
M/R \leq (1 + \sqrt{1 + 3 Q^2/R^2})^2/9 \, .
\end{equation}
Again the bound saturates for a thin charged shell with $2p_t = \rho_m$ \footnote{In Ref. \cite{and-ekl-rei}, numerical investigations have shown that there are two types of saturating solutions, one of thin shell while the other for which $M=Q=R$ corresponding to the extremal distribution, $M/R=1$ and $\alpha^2=1$.}. In contrast to the earlier expression (1), this is not a direct compact relation but rather an implicit relation where $M/R$ is given in terms of $Q^2$ and $R^2$ and not in terms of $\alpha^2$ alone. It however reduces to the Buchdahl limit for $Q=0$, and for the extremal case $\alpha^2 = 1$ to $M/R = 1$. In the extremal case, boundary of distribution
 coincides with horizon of charged black hole. This means compactness bound extends down to black hole horizon \footnote{In that case boundary of distribution, which is located at $p=0$, is null and hence pressure has to be a function of both $r$ and $t$. Therefore interior cannot remain static. Staticity of interior would pick up the former -- infinitely thin shell solution.}. Further it turns out that this limit also saturates  \cite{lemos14} for a charged analogue of of the Schwarzschild interior solution with $\rho_m + Q^2(r)/(8\pi r^4) = constant$ \cite{coop, florides} when $p_c(r=0)\to\infty$ \cite{guil, lemos10, lemos14a}. \\

Clearly the two limits are distinct and inequivalent and they have different bounds on charge an object could have. The extremal bound for the former \cite{roth}, $\alpha^2 = 9/8 \, >1$ is over-extremal for the latter \cite{andrea09}.  That is, the extremal limit for the former would be overcharged case for the latter. Unlike the Buchdahl limit for neutral object (which is the same for both prescriptions (1) and (2)), there are two distinct limits for charged object depending on interior distribution and its equation of state.   \\

In this paper, our main aim is not so much to discuss different compactness limits and their relative merits but instead it is to pose a question: {\it does the unique exterior solution describing field of neutral (Schwarzschild) and charged (Reissner - Nordstro$\ddot o$m) object have any say on its compactness} ? At the first sight, one would say that compactness should depend upon internal structure involving equation of state of fluid distribution and binding energy. It is the fluid properties that should determine how compact an object should be, and this is how the compactness bounds alluded above have been obtained. On the other hand gravity is universal and hence interior and exterior may be intimately related. It may not therefore be totally out of place to pose such a question. In the exterior, quantity that may have some bearing on the issue is gravitational field energy which can be evaluated anywhere outside the object. It is gravitational energy lying outside a sphere of radius $R$ and it monotonically goes to zero as $R\to\infty$. In the Newtonian limit it is given by $-M^2/2R$ similar to electrostatic energy, $Q^2/2R$. It is of course an extensive quantity and is always negative, however we shall henceforth refer to its absolute value all through \footnote{The effective gravitational mass of a charged object would be given by $M - Q^2/2R$ at any radius $R$ where electrostatic energy lying exterior to $R$ has been subtracted from $M$.}. Note that compactness ratio $M/R$ for a sphere of radius $R$ like gravitational field energy monotonically decreases to zero asymptotically. That is why compactness limit could translate into limit on gravitational field energy. It is therefore conceivable that compactness limit could be given in terms of gravitational field energy which is entirely determined by the unique exterior metric. In general relativity, energy and more so gravitational field energy is a very contentious issue. However in spherical symmetry there exist some physically reasonable prescriptions, and the one due to Brown and York \cite{bro-yor} is in particular generally accepted. It may be mentioned here that like positive mass theorem, its positivity has also been established \cite{liu-yau}. The Brown-York prescription gives total energy contained inside a sphere of given radius which is sum of non-gravitational matter energy and gravitational field energy. Hence by subtracting matter energy from that, one can obtain gravitational field energy lying outside. \\

It turns out that the Buchdahl compactness limit for an uncharged object is given by gravitational field energy being less than or equal to half of its non-gravitational matter energy. We propose that this should be true in general even when object is charged;i.e. {\it the Buchdahl compactness limit is defined by gravitational field energy being less than or equal to half of its non-gravitational matter energy in general for a static object whether charged or neutral.} With this definition, the limit we get for a charged object is that given in Eq (1). In fact it follows from $M$ in the neutral Buchdahl limit being replaced by effective gravitational mass, $M-Q^2/2R$, of charged object. In terms of surface potential, the limit reads for both charged and neutral object as $\Phi(r=R) = (M - Q^2/2R)/R \leq 4/9$. In higher dimensions, it has been strongly argued \cite{dad15} that one should consider pure Lovelock equation for gravity which includes only one $N$th order term, with no sum over lower orders. The linear $N=1$ is the Einstein while quadratic $N=2$ is the Gauss-Bonnet, and so on. Then following the Buchdahl procedure for fluid sphere in pure Lovelock gravity, the limit has been obtained \cite{sum-dad16} that reads as
\begin{equation}
(\frac{M}{R})^n \leq \frac{2N(d-N-1)}{(d-1)^2} \, \, \, ,  n = (d-2N-1)/N
\end{equation}
where $N$ is degree of Lovelock polynomial action. In particular it exactly agrees with the Buchdahl limit, $M/R \leq 4/9$ in $d=3N+1$ where pure Lovelock potential falls off as $1/R$ \cite{sum-dad18}. \\

We shall begin by a brief recall of the Brown-York definition of quasi-local energy for a charged object described by the unique Reissner - Nordstro$\ddot o$m metric. From that we would evaluate gravitational field energy. The compactness limit would then be given by {\it gravitational field energy being less than or equal to half of non-gravitational matter energy}. We shall conclude with a discussion. \\

\section{Gravitational field energy and Buchdahl limit}

Let us briefly recall the Brown-York prescription \cite{bro-yor} in which it is envisioned that a space-time region is bounded in a $3$-cylindrical timelike surface bounded at the two ends by a $2$-surface. Then Brown-York quasilocal energy is defined by
\begin{equation}
E_{BY}= \frac{1}{8\pi} \int{d^2x\sqrt{q}(k-k_0)}~,
\end{equation}
where $k$ and $q$ are respectively trace of extrinsic curvature and determinant of metric, $q_{ab}$ on $2$-surface. The reference extrinsic curvature, $k_0$ is of some reference space-time, which for asymptotically flat case would naturally be Minkowski flat. This is the measure of total energy contained inside a sphere of some radius $R$ around a static object.  The evaluation of the above integral for the Reissner - Nordstr{$\ddot o$}m metric yields,
\begin{equation}
E_{BY}(r\leq R) = R - \sqrt{R^2 - 2MR + Q^2},
\end{equation}
which at large $R$ approximates to
\begin{equation}
E_{BY} = M - (Q^2/2R - M^2/2R) = M - Q^2/2R + M^2/2R .
\end{equation}

This prescription envisions an infinitely dispersed distribution of bare ADM mass $M$ \cite{ADM} at infinity, while collapsing under its own gravity it picks up gravitational field energy as well as electrostatic field energy. Then gravitational field energy would be given by subtracting from it mass $M$ and electrostatic energy $Q^2/2R$. That is to evaluate gravitational field energy, subtract non-gravitational matter energy $E_m(r\leq R) = M - Q^2/2R$ from total energy, $E_{BY}(R)$ contained inside radius $R$. Thus gravitational field energy (its absolute value as alluded before) lying outside $R$ is given by
\begin{equation}
E_{GF}(r\geq R) = E_{BY} - (M - Q^2/2R) = R - \sqrt{R^2 - 2MR + Q^2} - (M - Q^2/2R).
\end{equation}
Note that total energy, $E_{BY}(r\leq R) = E_m(r\leq R) + E_{GF}(r\geq R)$. 
Note that it is the absolute value of gravitational field energy which is being slated against positive matter energy. \\

In Ref. \cite{dad97}, a remarkable and novel prescription for location of black hole horizon was proposed: {\it equipartition of non gravitational matter and gravitational field energy defines horizon radius of a static black hole};i.e. $E_m = E_{GF}$. This was further extended to pure Lovelock gravity in \cite{sum-dad15}. The natural question then arises, could we similarly also prescribe compactness limit ? Let's set $E_{GF} = \beta E_m$, and then ask what value of $\beta$ would give the Buchdahl limit $M/R \leq 4/9$? The answer turns out to be $\beta=1/2$ \cite{sum-dad16}. We thus propose that this should be true in general for any static object even when it is charged. Remarkably the limit it gives for a charged object is as given in Eq. (1) \cite{roth}. \\

Since gravitational field energy is determined entirely by the exterior unique solution, compactness limit could be found without reference to interior distribution, may what that be. This is a new prescription for compactness limit of a static star/object which is insightfully novel and interesting. It entirely depends upon the unique exterior space-time described by the Reissner - Nordstr{$\ddot o$}m (Schwarzschild) metric for a charged (neutral) object. By employing the Brown-York definition of quasilocal energy \cite{bro-yor}, it is possible to compute gravitational field energy of a static object lying outside certain radius $R$.  \\

The compactness limit is given by the equation $E_{GF} \leq (1/2) E_m$,
\begin{equation}
 R - \sqrt{R^2 - 2MR + Q^2} - (M - Q^2/2R) \leq \frac{1}{2} (M - Q^2/2R)
\end{equation}
which for $M-Q^2/2R \neq 0$ \footnote {When $M = Q^2/2R$, it defines the "classical electron radius" where mass is entirely due to electric field energy, and effective gravitational mass vanishes.} yields the Buchdahl analogue compactness limit for charged object
\begin{equation}
M/R \leq \frac{8/9}{1 + \sqrt{1 - 8\alpha^2/9}}~, \, \alpha^2 = Q^2/M^2~.
\end{equation}
Note that it could also be written as
\begin{equation}
\Phi(R) = (M-Q^2/2R)/R \leq 4/9~.
\end{equation}\\

This is exactly the limit obtained in \cite{roth} by considering absolute stability of interior consisting of uniform density neutral fluid enveloped by a thin charged shell. It reduces to the Buchdahl limit for $Q=0$, and it is, $M/R \leq 4/9, 2/3, \, 8/9$ for $\alpha^2 = 0, 1, \, 9/8$ respectively. Note that here extremality is not at $\alpha^2 =1$ but at $\alpha^2 = 9/8$ and $M/R < 1$ always. That means, no configuration can ever reach horizon and object always remains non-black hole. This is in contrast to the limit (2) \cite{andrea09} given above where $M/R=1$ for the extremal $\alpha^2 = 1$ case. The upper bound on charge a body could have is $\alpha^2 \leq 9/8$, which would be over-extremal for black hole. The most remarkable feature of this prescription is that it only refers to the exterior Reissner - Nordstr{$\ddot o$}m metric without any reference to interior at all, may what that be! \\

There is another prescription for gravitational field energy due to Lynden-Bell-Katz \cite{lyn-kat} where they define gravitational field energy density for the exterior Schwarzschild metric in isotropic coordinates. By transforming from isotropic to curvature coordinates and after much involved calculations, we have been able to establish equivalence between gravitational field energy computed from the two different prescriptions due to Brown-York and Lynden-Bell-Katz \cite{dad-che86} for a Reissner - Nordstr{$\ddot o$}m charged object. It is interesting that the two distinct prescriptions yield the same measure for gravitational field energy which is always fraught with ambiguity because there exists no covariant definition for it. What it indicates is the fact that for the  specific case of static spherically symmetric spacetime, it could indeed be evaluated in physically and intuitively satisfactory manner, and the two prescriptions in question give the same measure. \\

The Brown-York prescription could be extended to pure Lovelock gravity as well \cite{sum-dad15}, and the analogue of Eq. (5) would read as
\begin{equation}
E_{BY}(r\leq R) = R^n(1 - \sqrt{1 - 2(\bar M/R)^n}) \, \, \, , n = (d-2N-1)/N
\end{equation}
where $\bar M = M - Q^2/2R$ and $M^n$ is the ADM mass.  Then gravitational field energy would be given by
\begin{equation}
E_{GF}(r\geq R) = E_{BY} - \bar M^n = R^n(1 - \sqrt{1 - 2(\bar M/R)^n}) - \bar M^n \, .
\end{equation}

Clearly black hole horizon is defined when $E_{GF} = \bar M^n$, and the compactness limit for a static object, $E_{GF} = 1/2 \bar M^n$, would be given by
\begin{equation}
\Phi(R) = (\bar M/R)^n \leq 4/9 \, .
\end{equation}
This however does not agree in general with the one obtained in Eq. (3) following the Buchdahl prescription unless $d=3N+1$. That is, in pure Lovelock gravity the prescription of compactness limit, gravitational field energy being less than or equal to half of non-gravitational matter energy, does not agree in general with the one obtained from the Buchdahl procedure unless $d=3N+1$.


\section{Discussion}

It is interesting that equality of gravitational field and non-gravitational matter energy defines black hole horizon while the former being half of the latter defines the Buchdahl compactness limit. In \cite{dad97} it was argued that timelike particles experience gravitational acceleration which is caused by non-gravitational matter energy while photons experience no acceleration but instead they only experience curvature of space which is caused by gravitational field energy \cite{dad12}. As horizon is approached timelike particle tends to photon $v\to c$, and hence at the horizon the measures of their respective sources (non-gravitational matter energy producing acceleration for timelike and gravitational field energy producing space curvature for null) should be equal. That is how equipartition of gravitational and non-gravitational energy defines location of horizon. This was our motivation for seeking compactness limit similarly, but how do we understand it physically? Secondly how does exterior metric by itself without reference to interior distribution at all determine compactness limit?  \\

The relevant quantity for compactness is binding energy. How do we define binding energy, total energy contained inside boundary of star minus matter energy, and it is computed entirely from interior distribution without reference to exterior solution? This is exactly also the definition of gravitational field energy lying outside certain radius $R$, and it could be computed by using only the exterior metric without reference to the interior metric or distribution. The former however lies in interior and could be computed on boundary of object or inside without reference to exterior metric, while the latter is entirely in exterior and hence could be computed at anywhere outside. The two would however coincide when they are evaluated at the boundary. At the boundary measure of binding energy is the same as that of gravitational field energy lying outside the object. Gravitational field energy and binding energy could be looked upon as reflection of each-other with both coinciding at the boundary. This is how condition on gravitational field energy determines compactness limit. Thus there is intimate connection between interior and exterior. What it says is that an object could be as compact as it wishes so long as its gravitational or binding energy does not exceed half of its non-gravitational matter energy. The remarkable point is that it is computed entirely using the exterior metric which is unique for a static object whether neutral or charged. On the other hand binding energy depends on interior fluid distribution and its structure in terms of equation of state, and there would thus exist different solutions depending upon fluid  properties. \\


The question however remains why should it be half of matter energy to give compactness limit? This reminds of the Virial theorem for a gravitationally stable equilibrium configuration where averaged kinetic energy is half of potential energy. Here for the compactness limit we have gravitational field energy being less than or equal to half of non-gravitational matter energy. All what one can say is that definition of relativistic kinetic energy, total energy minus rest mass energy, is quite similar to that of gravitational field energy, total energy contained inside some radius $R$ minus non-gravitational matter energy. Further setting for the Virial theorem is cluster of gravitating bodies. The question arises how could it be applied to a fluid interior of a static object? If it could be, it would mean fluid elements behave like a system of gravitating bodies. Perhaps one should formulate kinetic theory to gain further understanding, all what could be said is that this raises a very interesting and insightful question. \\

In the case of neutral object there exists the unique Schwarzschild interior solution for the stiffest equation of state of uniform density fluid distribution. The compactness limit follows from the requirement that pressure at the center, $p_c(r=0) \leq \infty$. Here the limit is saturated for $p_c(r=0)\to\infty$. Unfortunately for the charged case, there exists no such unique solution. We may however say that constant density neutral fluid being enveloped by a thin charged shell \cite{roth} perhaps comes closest, and that yields the Buchdahl limit (1). Note that our prescription in terms of gravitational field energy also gives the same limit as given in Eq. (9). \\

Gravitationally electric charge contributes repulsively and hence a charged object would be less compact than the neutral one. This is because effective active gravitational mass goes as $\bar M = M - Q^2/2r$ and gravitational potential as $\Phi = \bar M/r$. In terms of the surface potential, the Buchdahl compactness limit is given by $\Phi(R) \leq 4/9$ always irrespective of object being charged or neutral. This is because decrease in effective active gravitational mass is exactly compensated by decrease in radius. The compactness bound $\Phi(R) \leq 4/9$ also translates into a bound on surface red-shift, $Z_s \leq 3$ and on escape velocity, $v^2 = 2\Phi(R) \leq 8/9$. This would indicate that a non black hole object would always have surface red-shift $Z_s \leq 3$. This is a definitive observational prediction. However it may be very difficult to measure red-shift of individual star. \\

There exists an upper bound on charge, $Q^2/M^2 \leq 9/8$, a body could have. Clearly unlike black hole there is nothing that prohibits a body from having $Q^2 > M^2$ because $R^2 - 2MR +Q^2 > 0 $ and $M/R<1$ always. A charged compact body could exist with charge exceeding its mass. Thus there is a bound for maximum charge not only for black hole but also for a non black hole object which is interestingly greater than that for the former. It is not for the first time one is encountering this bound, $\alpha^2 \leq 9/8$, it appears in photon sphere radius for the Reissner - Nordstr{$\ddot o$}m metric, given by
\begin{equation}
r_{ph\pm} = \frac{3M}{2}(1 \pm \sqrt{1 - 8\alpha^2/9})
\end{equation}
requiring $\alpha^2 \leq 9/8$. What it means is that when $\alpha^2 >9/8$, there exists no photon circular orbit. That is space curvature is not strong enough to make photon geodesic close on itself to admit a circular orbit. It should be noted that outer and inner radii are respectively $r= 3M, 0$ for uncharged object while they are $ r = 2M, M$ for extremal charged black hole, $\alpha^2 = 1$ \footnote {This is quite in contrast to the Kerr black hole where both merge into outer horizon for extremal black hole.}. The two merge for extremally charged object for $\alpha^2 = 9/8$. This bound on charge is also discussed in a recent paper \cite{hod} considering upper bound on gravitational mass of stable spatially regular charged compact objects obeying the limit (2). There may perhaps exist some reconciling relation between the two compactness limits. It would be really interesting to unravel that. Another interesting question that arises is to find an interior solution for a charged object with $1 < \alpha^2 \leq 9/8$ so as to prove that such distributions indeed exist. It may be difficult to find analytic solutions, one may have to resort to numerical simulations. \\


One might ask how about a rotating object which is astrophysically most interesting and relevant? It is well-known that rotation brings in many complications and technical difficulties. We have the well-known axially symmetric stationary Kerr solution of the Einstein vacuum equation which truly describes only a rotating black hole and {\it not a rotating body}. A rapidly rotating object cannot have exact axial symmetry because of flattening at the poles giving rise to multipole moments and it cannot therefore have spherical topology as is the case for Kerr black hole horizon. Since  black hole can have no hair, all moments have evaporated away before horizon is formed. For interior, there is well known solution under the slow rotation approximation \cite{har-tho}. There is an extensive study of the problem \cite{neu+} and it is shown that numerical solutions exist for rotating fluid as well as kinematic matter sources. Also noteworthy is a very recent numerical simulations for constructing models of dynamically stable ergostars \cite{sha+}. Since the Kerr solution is for a rotating black hole and not for a rotating object, one has also to construct exterior solutions for these interior solutions. For non black hole rotating object exterior solution would not perhaps be unique. For rotating object the problem is thus much more involved and complicated.  \\

Finally our main aim in this brief note is to draw attention to a very novel and insightful prescription for Buchdahl compactness limit for a static object involving gravitational field energy, and thereby the limit is obtained entirely from the unique exterior solution with no reference to interior at all, may what that be. The limit so obtained may or may not be entirely satisfactory yet it is interesting and insightful. It is remarkable that it is given by in terms of gravitational potential, $\Phi(R) \leq 4/9$ irrespective of object being charged or neutral. \\

Acknowledgment: I wish to warmly thank Parampreet Singh for suggesting consideration of the Lynden-Bell-Katz energy and V. Chellathurai for help in verifying it gives the same value as that due the Brown-York  prescription.  \\

\end{document}